\documentclass[useAMS,usenatbib,usegraphicx]{mn2e}
\usepackage{fixltx2e} 
\usepackage{url}    
\usepackage[modulo]{lineno} 
\usepackage[svgnames]{xcolor}  %
\newcommand{\arcdeg}{\mbox{$^\circ$}}
\newcommand{\uv}{\mbox{$u$-$v$}}

\newcommand{\Msol}{\mbox{M\raisebox{-.6ex}{$\odot$}}}

\newcommand{\pyear}{\mbox{\% yr$^{-1}$}}
\newcommand{\pyrsq}{\mbox{\% yr$^{-2}$}}

\newcommand{\kms}{\mbox{km s$^{-1}$}}

\newcommand{\PSRB}{\mbox{PSR B0531+21}}
\newcommand{\RPWN}{\mbox{$R_{\mathrm{PWN}}$}}
\newcommand{\RFIL}{\mbox{$R_{\mathrm{fil}}$}}
\newcommand{\Rexp}{\mbox{${\scriptstyle R}_{\rm exp}$}}
\newcommand{\AD}{\mbox{\textsc{ce}}}
\newcommand{\Ra}[4]{\mbox{${#1}^{\mathrm h} \; {#2}^{\mathrm m} \; {#3}\fs{#4} $}}
\newcommand{\dec}[4]{\mbox{${#1}\degr \; {#2}\arcmin \; {#3}\farcs{#4} $}}

\newcommand{\lesssim}{\mbox{\raisebox{-0.3em}{$\stackrel{\textstyle <}{\sim}$}}}

\newcommand{\tablenotemark}[1]{$^{\mathrm #1}$}
\newcommand{\tablenotetext}[2]{\noindent$^{\mathrm #1}$ #2\\}
\newcommand{\phn}{\phantom{1}}

\title[New expansion rate measurements of the Crab Nebula]{New
  expansion rate measurements of the Crab Nebula in radio and optical}
\author[Bietenholz \& Nugent]{M. F. Bietenholz$^{1,2}$, and
Nugent, R. L.$^3$ \\
$^1$Department of Physics and Astronomy, York University, Toronto,
M3J~1P3, Ontario, Canada \\
$^2$Hartebeesthoek Radio Observatory, PO Box 443, Krugersdorp,
1740, South Africa \\
$^3$IOTA, International Occultation Timing Association, USA \\
}

\begin{document}
\date{\today;  {\em Accepted to MNRAS}}
 
\maketitle
\label{firstpage}

\begin{abstract}
We present new radio measurements of the expansion rate of the Crab
nebula's synchrotron nebula over a $\sim$30-yr period.  We find a
convergence date for the radio synchrotron nebula of \AD\ $1255\pm27$.
We also re-evaluated the expansion rate of the optical line emitting
filaments, and we show that the traditional estimates of their
convergence dates are slightly biased.  Using an un-biased Bayesian
analysis, \textcolor{black}{we} find a convergence date for the
filaments of \AD\ $1091 \pm 34$\@ ($\sim 40$~yr earlier than previous
estimates).  Our results show that both the synchrotron nebula and the
optical line-emitting filaments have been accelerated since the
explosion in \AD\ 1054, but that the synchrotron nebula has been
\textcolor{black}{relatively strongly accelerated}, while the optical
filaments have been \textcolor{black}{only slightly accelerated}.  The
finding that the synchrotron emission expands more rapidly than the
filaments supports the picture that the latter are the result of the
Rayleigh-Taylor instability at the interface between the pulsar-wind
nebula and the surrounding freely-expanding supernova ejecta, and
rules out models where the pulsar wind bubble is interacting directly
with the pre-supernova wind of the Crab's progenitor.
\end{abstract}


\section{Introduction}
\label{sintro}

The Crab Nebula is one of the most intensely studied objects in
astrophysics, yet it retains the power to surprise us \citep[see][for
  recent reviews]{BuehlerB2014, Hester2008}.  It is the remnant of a
supernova explosion in the year \AD\ 1054, which was witnessed by
Chinese and other astronomers \citep{StephensonG2003}.  The presently
visible nebula is bright at all observable wavelengths, and contains
one of the first known pulsars, \PSRB.  The pulsar's spin frequency is
30~Hz, and it is slowing down at a rate of $-3.78 \times
10^{-10}$~Hz~s$^{-1}$ \citep{Lyne+2015}.  The Crab Nebula is the
prototype of a pulsar-powered nebula, commonly known as a pulsar wind
nebula (PWN), where the rotational energy lost by the spinning down of
the pulsar powers the nebula\textcolor{black}{.} The energy input from
the pulsar, which emerges in the form of a wind of magnetic field and
relativistic particles, inflates a large bubble of relativistic fluid,
which emits synchrotron radiation.  The synchrotron-emitting fluid
expands into the supernova ejecta, with which it interacts both
dynamically and by photoionizing them.  The synchrotron-emitting
bubble and the optical-line emitting filaments of photoionized thermal
gas constitute the bulk of the presently visible nebula.

The Crab has recently been discovered to produce substantial flares
\textcolor{black}{at gamma-ray wavelengths}, where the emission at energies
$>100$~MeV increases by more than a factor of two on timescales of
days \citep[e.g.][and references therein]{BuehlerB2014}. The origin of
these flares is not yet well understood.  Since they are poorly
localized in gamma-rays, we obtained radio observations of the Crab
following a gamma-ray flare in 2012 August to look for a radio
counterpart.  Our initial results are reported in \citet{Crab2015a}:
we did not find any such radio counterparts.  We did, however, obtain
high-quality radio images of the Crab.  In the present paper, we use
those radio images (from epoch 2012), and compare them to earlier
radio images to more accurately determine the expansion rate of the
synchrotron nebula.

The PWN inflated by the pulsar continues to expand, and the Crab
nebula is young and close enough that the expansion has been directly
observed at different wavelengths.  In fact, not only is the Crab
expanding, it has been accelerated {\em since}\/ the \AD\ 1054
explosion.  Several authors have measured the proper
motion\textcolor{black}{s} of the optical filaments over the years
\citep[e.g.,][]{Duncan1939, Woltjer1958, Trimble1968, WyckoffM1977,
  Nugent1998}, and from these calculated a convergence date, which is
the date on which the filaments would \textcolor{black}{have been}
closest together if their present positions and proper motions are
extrapolated backwards in time.  This convergence date is later than
the known explosion date of \AD\ 1054, indicating an accelerated
expansion.

The expansion rate of the nebula \textcolor{black}{was also} determined
in the radio by \citet{Crab-expand}, who used VLA observations
\textcolor{black}{taken} between 1981 and 1988\@.  This radio
measurement gave a determination of the expansion speed of the pulsar
wind bubble (as opposed to that of the optical line-emitting
filaments), and showed that the \textcolor{black}{former} was expanding
at a rate slightly higher than, but within the uncertainties, of that
observed for the filaments.

In the canonical picture of the Crab, such acceleration is in fact
expected, because the PWN is expanding {\em into} \textcolor{black}{the}
supernova ejecta, which themselves are still freely expanding.  Since
the ejecta expand with velocity proportional to the radius, as the PWN
expands into the ejecta, it must move ever faster to overtake them, so
as long as there is continued energy input from the pulsar and the PWN
remains within the freely-expanding ejecta, the PWN bubble is expected
to accelerate \citep[see, e.g.][]{ReynoldsC1984, Chevalier1984-PWN}.

If SN 1054 was a normal supernova, then it would be expected to have
released about $10^{51}$~erg of energy.  The presently visible Crab
nebula, however, contains at most 10\% of this energy.
It is mostly thought that the remainder of the expected $10^{51}$~erg
resides in freely expanding ejecta, outside the presently visible
nebula.  \textcolor{black}{There has as yet been no convincing direct
  detection of this ``outer shell'' of ejecta, despite the fact that
  it is predicted to be substantially ionized by the flux from the PWN
  \citep{LundqvistFC1986, Wang+2013}}.

Nonetheless, circumstantial evidence does suggest that the presently
visible nebula is confined by the unseen massive ejecta.  In this
picture, the optically-bright filaments are the product of a
Rayleigh-Taylor instability between the low-density but high-pressure
synchrotron-emitting fluid and the massive ejecta.  \citet{Hester2008}
gives a summary of the arguments for this interpretation.

We note here, however, that there is an alternate interpretation,
which is that SN~1054 was a low-energy electron-capture event, with
only a small ejected mass, and that the presently visible filaments
are the result of the PWN interacting with the circumstellar material
\citep{Smith2013, TominagaBN2013, Moriya+2014}.
The progenitor in this case is expected to have been a
super-asymptotic giant branch star, in other words one at the upper
end of the mass range of the asymptotic giant branch, with a mass of
around 8 to 10 \Msol, although the \textcolor{black}{mass is} not well
constrained \citep[see][and references therein]{Moriya+2014}.  Such
progenitors are expected to produce slow, dense winds before the SN
explosion.

If the Crab PWN is currently expanding into the freely expanding SN
ejecta, then it should, as mentioned above, be the case that the
synchrotron bubble is expanding more rapidly than the optical
filaments.  The measurements of \citet{Crab-expand}
\textcolor{black}{already suggested} that this might be the case, but a
better measurement of the expansion of the synchrotron bubble is
required to be sure.  Although the synchrotron emission is visible in
the optical and even in the X-ray, such a measurement is most easily
done in the radio where the synchrotron bubble is most clearly
visible.  We therefore undertook a new and more accurate determination
of the expansion rate of the Crab's PWN using radio images, as well as
\textcolor{black}{a re-evaluation of} the expansion rate of the optical
line-emitting filaments determined from proper motion
measurements. The radio and optical data allow us to measure the
expansion rate of two different components of the nebula, namely the
low-density but high-pressure PWN and the massive line-emitting
filaments, respectively.

\section{Observations and data reduction}
\subsection{VLA observations}
\label{svlaobs}

For the purposes of measuring the expansion rate of the Crab Nebula,
we assemble a collection of 5-GHz images taken between 1981 and 2012.
Images made from data taken at a specific time, and thus only using a
single VLA array configuration\footnote{The VLA array configurations
changes approximately three times per year.}
would be best-suited for determining the
expansion.  We concentrate therefore on using images made only from
B-configuration data, which was used for the observations between 1998
and 2012\footnote{Note that we do not use images deconvolved with a
  default image constructed from earlier observations as was done in
  \citet{Crab2015a, Crab-2004}, since that would bias the measured
  expansion rate to that used to construct the default.}.
However, due to various reasons, for some of the earlier epochs we use
images made from combinations of several configurations, spatially
filtered so as to approximate the B-configuration-only images.  We
will note these epochs below.

\textcolor{black}{Using observations made only} in a single VLA array
configuration is the equivalent of applying a spatial filter to the
images, limiting both the highest spatial frequency (highest
resolution) and the lowest one (largest recoverable size).  In
particular, for the B-array at 5-GHz, the resolution is limited to
$\sim 1\farcs 2$ and the largest recoverable structure is
$\sim$30\arcsec. Although the range of spatial frequencies sampled is
similar in all our epochs it will not be identical due to slightly
different \textcolor{black}{observing} frequencies, different observing
hour angles and the occasional failure of an antenna.  In order to
unify our images as much as possible, we further spatially filter our
images using a Gaussian kernel.  First, we use a spatial low-pass
filter to reduce the effective resolution to 2\farcs0 $\times$
1\farcs8 at p.a.\ 80\arcdeg, and secondly a high-pass filter with FWHM
of 20\arcsec\ to exclude poorly sampled large-scale structure.

As already mentioned, we use the 2012 radio observations described in
\citet{Crab2015a}.  \textcolor{black}{We give here only a brief
  description of the observations:} there were two sessions of VLA
observations on 2012 August 20 and 26, spaced 6 days apart (observing
code 12A-486).
We used a bandwidth 2048 MHz around a central frequency of 5567 MHz,
with a total of 5 hours per session.  The array was in the B
configuration, resulting in native resolutions of
$\sim$1\arcsec\ FWHM, but as mentioned we low-pass filtered them to
reduce the effective resolution to 2\farcs0 $\times$ 1\farcs8 for our
expansion measurement.


We also used data from 2000 Feb.\ 11 and 2001 Apr.\ 17, described in
\citet{Crab-2004}.  These observations were taken with centre
frequencies of 4.885 and 4.625 GHz.  Both frequencies were combined
for imaging.  We further used data from 1998 August 8, which is
described in \citet{Crab-2001}, and which was taken
\textcolor{black}{with} the same centre frequencies.

We further used data taken between 1987 May to 1988 March, again taken
at 4.885 and 4.625 GHz and described in \citet{Crab-jet},
\citet{Crab-Faraday}, \citet{Crab-expand} and \citet{Crabwisp-1992}.
For these epochs, the calibrated visibility data were no longer
available, and we \textcolor{black}{worked} with the deconvolved images,
which were made with the \textcolor{black}{combined} data from the B, CD
and D configurations \textcolor{black}{using} maximum entropy
deconvolution.  We again high-pass filter them with a Gaussian of FWHM
20\arcsec.  This filtering will remove almost all of the information
contributed by the D array, and some of that contributed by the
C-array.  We take as the effective date of this image therefore a
weighted average of those of the B and C configurations only, with the
B configuration given double the weight of the C configuration.  Since
all three configurations were observed within one year the effective
date is not overly sensitive to variations in this weighting and our
results below do not \textcolor{black}{depend on it.}

Finally, we also used data from B-configuration taken on 1982
Oct.\ 16.  \textcolor{black}{Since the signal-to-noise ratio of the
  B-configuration data alone was low, we added some data from the C
  configuration,} taken on 1981 Nov.\ 29.  \textcolor{black}{Both sets
  of observations were at 4.9~GHz and had a bandwidth of 12.5~MHz}.
These observations are described in \citet{WilsonSH1985b} and
\citet{Crab-expand}.  We re-reduced the archival visibility
data\textcolor{black}{, combined the data from both configurations, and
  then} made an image using CLEAN deconvolution.  We apply the same
prescription as we used for the 1987 epoch to determine the effective
date, of taking a weighted mean of the B and C configuration observing
dates, with the B-configuration being given twice the weight,
\textcolor{black}{and again high- and low-pass filtered the resulting
  image as we did the others.}

\section{Expansion of the synchrotron bubble}
\label{sexp}

\begin{figure*}
\centering
\includegraphics[width=0.90\textwidth]{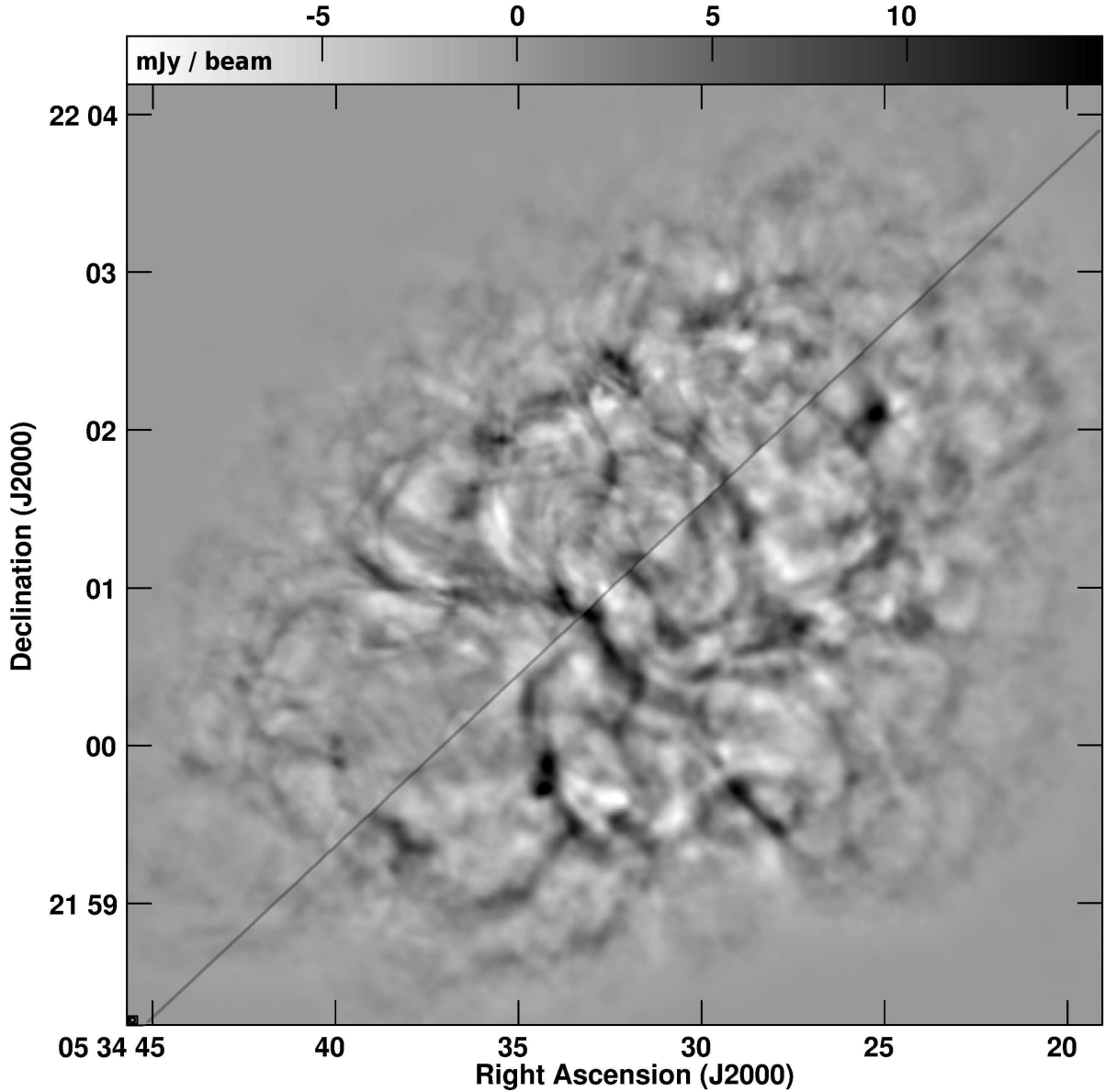}
\caption{A composite image to illustrate the expansion of the Crab
  synchrotron nebula.  On the top left is the image from 2012
  Aug.\ 26, while on the lower right is the one from 1987 Dec.  Both
  images are 5-GHz VLA radio images, which have been spatially
  band-limited to show only emission on \textcolor{black}{a limited
    range of} angular scales by convolving to limit the resolution to
  2\farcs0 $\times$ 1\farcs8 at p.a.\ 80\arcdeg\ (FWHM), and high-pass
  filtering at 20\arcsec\ (FWHM). Note that the image brightness can
  have negative as well as positive values because of the high-pass
  filtering.  See text for a fuller description of the data and the
  processing of the images.}
\label{fimgcomp}
\end{figure*}

We first measure the expansion rate of the synchrotron bubble using
the radio images.  To illustrate the expansion we show a composite of
the images from 1987 and 2012 in Figure~\ref{fimgcomp}.  Since there
are few sharply-defined features from which the expansion could be
directly measured, we use the same approach \textcolor{black}{for
  determining} the expansion as was used in \citet{3C58-2006},
\citet{G21.5expand} and \citet{Crab-expand}, \textcolor{black}{and which
  was} originally developed by \citet{TanG1985}.  We repeat a brief
description here for the convenience of the reader: rather than
determining the proper motion of individual features, we measure the
overall expansion \textcolor{black}{by determining} the scaling between
a pair of images by least-squares. This is accomplished by using the
MIRIAD \citep{SaultTW1995} task IMDIFF
which determines how to make one image most closely resemble another,
by calculating unbiased estimators for the scaling in size, $e$, the
scaling and the offset in flux density, $A$ and $b$ respectively, and
the offsets in RA and decl., $x$ and $y$ respectively, by least
squares.  Our chief interest is in the expansion factor, $e$, but
because of uncertainties in flux calibration, absolute position, and
image zero-point offsets caused by missing short spacings and
self-calibration and slightly different observing frequencies, all
five parameters needed to be determined.

We calculate the expansion from images which were {\em not} corrected
for the primary beam response.  Although the attenuation due to the
primary beam is appreciable towards the edge of the nebula at
5~GHz\footnote{At the approximate radius of the Crab, 180\arcsec, from
  the pointing centre and 5.0 GHz, the measured brightness 74\% of the
  true value because of the falloff in the primary beam
  response.}\textcolor{black}{, the} uncorrected images are preferred
for the expansion calculation because the noise is uniform over the
image, which is a desirable property for an algorithm that computes a
least-squares fit over the whole image.  The use of the uncorrected
images should not affect our expansion results because the frequency
of observation, and thus the primary beam response, was almost the
same at all our epochs and the expansion is small.
We re-imaged or re re-sampled all the earlier observations to have the
same pixel spacing (0.27\arcsec) as the 2012 ones, and also, if
required, J2000 coordinates.  For each pair of images we compute $e$.
To reduce biases, we also repeated the calculation but inverted the
order of the images, which causes $1/e$ to be determined.  As our
final value of $e$ was the average of the two runs (although in all
cases the two values of $e$ were consistent to within 0.001).  The
uncertainty in $e$ is difficult to estimate, we conservatively adopt a
value of 0.002 (however, we show below that the scatter in the derived
expansion rate suggests a lower uncertainty in $e$ of $\sim 0.0009$).
We give our values of $e$ for various pairs of images in
Table~\ref{texp}.  We define the fractional expansion rate,
\Rexp\textcolor{black}{,} as the percentage increase in size of the
nebula per year.  The weighted mean value of \Rexp\ over the period of
1982 to 2012, was $0.134 \pm 0.005$~\pyear\ (at the weighted-mean
epoch of 1996.6).

From each pair of images, we can also calculate a ``convergence date''
for the nebula under the assumption of constant-velocity expansion
from a single origin.  Let $t_1$ and $t_2$ be the
\textcolor{black}{times of two} images measured with respect to the time
of the explosion, and $\Delta t = t_2 - t_1$.  For constant velocity
expansion, the value of $e$ between $t_1$ and $t_2$ is just $(t_1 +
\Delta t)/t_1$.  From any \textcolor{black}{determination} of $e$ we can
therefore calculate the value of $t_1$ and thus estimate the explosion
date.  Since the assumption of constant-velocity expansion is well
known not to hold for the Crab, we term this date the ``convergence
date'' \textcolor{black}{or convergence year}, which in the Crab's case
will be somewhat later than the \textcolor{black}{actual explosion year}
of \AD\ 1054.

The weighted mean convergence year calculated from our measurements in
Table~\ref{texp} is \AD\ $1243.4 \pm 9.3$, with the uncertainty
obtained by scaling the input uncertainties to obtain $\chi^2_8 = 8$.
We also performed a Bayesian calculation to estimate both the
convergence year and the true uncertainties in $e$, which as mentioned
above were not reliably estimated {\em a priori}.  We used a uniform
prior distribution for the convergence year and a Jeffrey's prior
($p(\sigma) = 1/\sigma$) for the uncertainties in $e$.
The posterior distribution of the convergence year was obtained
through Markov Chain Monte-Carlo (MCMC) sampling\footnote{We used the
  PyMC package, version 2.3, by C. Fonnesbeck, A. Patil, D. Huard, and
  J. Salvatier to implement the Markov Chain Monte-Carlo calculation.
  This package is available at
  \url{http://pymc-devs.github.io/pymc/README.html}.}.  We obtained a
consistent value of \AD\ $1240 \pm 12$, with the estimate of the
measurement uncertainties in $e$ being 0.0009\textcolor{black}{.}

Is there any bias in this determination of the convergence date?  We
explored the possibility of a bias as follows: we made a copy of a
radio image, and expanded it by a factor of exactly 1.02 (i.e.\ by
2\%), then added noise.  For the noise, we used Gaussian random noise
with the value in each pixel being independent, then convolved with
the same CLEAN beam ($2\farcs0 \times 1\farcs8$ at p.a.\ 80\arcdeg) as
our images, so that our added noise has same spatial auto-correlation
function as that in the real images.  We then again used IMDIFF to
determine $e$ between this modified copy and the original image.  In
the absence of any bias, we expect to obtain $e = 1.0200$.  Over $n=8$
trials we found that the derived average value of $e$ was $1.0205 \pm
0.00006$, suggesting that any bias in $e$ is less than 0.00006.  We
therefore believe that our estimates of $e$ are not significantly
biased by the presence of noise in the images.

\begin{figure}
\centering
\includegraphics[width=0.48\textwidth]{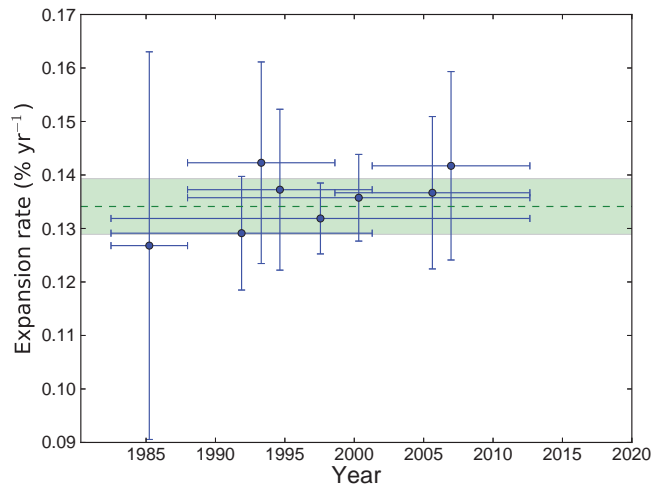}
\caption{The fractional expansion rate of the Crab Nebula (\Rexp), as
  calculated from pairs of radio images.  Each blue point shows a
  value of the expansion rate derived from one pair of images,
  with the horizontal bar showing the time interval over which the
  corresponding rate was calculated, and the vertical one the
  uncertainty in the derived expansion rate.  The green dashed line
  shows the weighted mean of \Rexp\ of $0.134 \pm 0.015$~\pyear, with
  the shaded area showing the $1\sigma$ uncertainty on the mean.  Near
  the outside edge of the nebula, \Rexp = 0.135~\pyear\ corresponds to
  2310~\kms\ ($\theta = 3\arcmin, D = 2$~kpc).}
\label{fexprate}
\end{figure}

Does the expansion rate of the Crab change with time?  We performed a
weighted fit to our measured values of \Rexp, taking the time of each
rate measurement as being the midpoint of the two epochs involved.  We
plot the resulting values of \Rexp\ in Figure \ref{fexprate}.  We
found that $\Rexp\ = +(5 \pm 9) \times 10^{-4} \, t_{1054} -(0.341 \pm
0.004)$~\pyear.
Note that \Rexp\ represents the fractional expansion rate, so for
constant velocity expansion since \AD\ 1054, $d\Rexp/dt$ would be
$-1.1 \times 10^{-4}$~\pyrsq, and for constant velocity expansion
since the convergence date of \AD\ 1250, it would be $-1.8 \times
10^{-4}$~\pyrsq.  
Although our results suggest that \Rexp\ is
increasing with time, they are well compatible with the expected
decrease with time (to within $1\sigma$), and we do not consider the
increase with time of \Rexp\ significant (see Fig.\ \ref{fexprate}).

\begin{table}
\caption{Images used for the expansion calculation}
\label{tepochs}
\begin{tabular}{l l l c c l l}
Date of observations & Julian date & \\
\hline
1982 Apr 24\tablenotemark{a} &
              2445084    \\
1987 Dec 29 & 2447159  \tablenotemark{b}  \\ 
1998 Aug 09 & 2451035 \\ 
2001 Apr 17 & 2452016 \\
2012 Aug 26 & 2456165 \\ 
\hline
\end{tabular} 
\\ \tablenotetext{a}{For this epoch, to increase the signal-to-noise,
  we combined data taken in the B and C configurations on 1982
  Oct.\ 16 and 1981 Oct.\ 29, respectively, but used data only at
  \uv~distance $> 3.3$~K$\lambda$.  As the time of these observations,
  we take a weighted mean, with the B-array, which contributes most of
  the information, being given double the weight of the C-array.  The
  resulting date is 1982 June 22.}
\tablenotetext{b}{For this epoch, we used an image made from B,CD and D
  configurations, taken on 1987 Nov.\ 21, 1988 Mar.\ 14 and 1987 May
  26, respectively.  Due to the spatial high-pass filtering which will
  isolate chiefly the information contributed by the B-array we use,
  as we did for the 1982 epoch, a weighted combination of the B and C
  configuration dates, with again the B-configuration being given
  twice the weight.  The resulting date is 1987 Oct.\ 31.}
\end{table}

\begin{table}
\caption{Expansion Factors}
\label{texp}
\begin{tabular}{l l l c c l l}
First Epoch & Second Epoch   & Interval & Expansion Factor \\
\hline
     &      &  (yr) &        \\
 1982 & 1987 &\phn5.52 &  1.007 \\ %
 1982 & 2001 & 18.82 &  1.024 \\ %
 1982 & 2012 & 30.18 &  1.040 \\ %
 1987 & 1998 & 10.61 &  1.015 \\ %
 1987 & 2001 & 13.30 &  1.018 \\ %
 1987 & 2012 & 24.66 &  1.034 \\ %
 1998 & 2012 & 14.05 &  1.019 \\ %
 2001 & 2012 & 11.36 &  1.016 \\ %
\hline
\end{tabular}
\end{table}

\section{Expansion of the line-emitting filaments}
\label{sexpfilaments}

We also re-examined the expansion rate as determined from the proper
motions of the line-emitting filaments observed in the optical.  We
\textcolor{black}{used} the proper motion measurements made by
\citet{Nugent1998}, on the basis of four published high-resolution
optical images of the Crab, with the first made in 1939 and the last
in 1992 \citep{Baade1942, Gingerich1977, Parker1995, WainscoatK1997}.
The details of the astrometrical reduction and proper motion
measurements are given in \citet{Nugent1998}.

Previously, the expansion age of the Crab had been determined by
extrapolating the present positions backwards using the (presently
measured) proper motions \citep[e.g.][]{Nugent1998}, and taking the
convergence date as the time at which the scatter amongst the
extrapolated positions was smallest, and the corresponding convergence
position as the mean position at that time.  This procedure, however,
does not properly take into account the uncertainties in the proper
motion measurements, and leads to a slightly, but significantly,
biased value for the convergence date, as we will illustrate below.

We will assume, like others workers \citep[e.g.][]{Duncan1939}, that
the filaments originated at some particular point in space,
\textcolor{black}{namely} the convergence point, at some particular
time, \textcolor{black}{namely} the convergence time, and that they have
moved at a constant velocity since then.  Since we know in fact that
the supernova was in 1054 \AD, a convergence date later than 1054
\AD\ implies that the filaments have not in fact been moving at
constant speed, but rather have been accelerated since 1054 \AD. A
more straightforward analysis would be to fix the explosion date at
its known value and more directly estimate the amount of acceleration.
However, we chose the more indirect method of estimating the
convergence date for easier comparison with earlier results.  The
amount by which the convergence date is later than 1054 \AD\ indicates
the amount of acceleration that has taken place.

The bias in the convergence date can easily be illustrated with the
following example.  Imagine that the Nebula were static, in other
words that the convergence date was infinitely far in the past (or the
future).  The true proper motions would be 0.  The measured values
would be randomly distributed about zero because of measurement
errors.  The smallest scatter in the extrapolated positions would be
near the present time, since no matter what the random motions, over
any length of time they are more likely to move the filaments
\textcolor{black}{farther} apart than closer together.  The date at which
the extrapolated position scatter is smallest would therefore be near
the present, not infinitely far in the past where the true convergence
date is.  This bias occurs even if the real proper motions are not
zero, in the sense that the date of smallest extrapolated position
scatter is always somewhat nearer to the present than the true
convergence date.

To take proper account of the errors in both the measured positions
and proper motions, we again turn to a Bayesian analysis.  We retain
the hypothesis that the proper motions are constant in time, and that
the filaments all originated at the convergence point in space and at
the convergence date in time, and it is the convergence point and
\textcolor{black}{date} that we wish to estimate.  For any given
convergence position and date, the present true filament positions and
proper motions are functionally related.
We treat the measurements of the positions and proper motions as
independent, and in our Bayesian analysis we also estimate the true
present positions of the filaments.  We take the measured proper
motions and positions to be Gaussian-distributed about the true
values, with standard deviations given by the observational
uncertainties.

We take the following prior distributions: uniform with a range of
\AD\ 500 to 1500 for the convergence year, Gaussian with $\sigma =
200\arcsec$ centred around the present mean position of the filaments
for the convergence position, and Gaussian with a with $\sigma =
10\arcsec$ centred on the measured positions for the true filament
positions.  Note that the measured filament positions are very
unlikely to be in error by 10\arcsec, so our prior distribution should
be minimally informative and should therefore have very little effect
on the derived posterior distributions.

We again turn to Markov Chain Monte-Carlo to estimate the posterior
distributions.  We obtain a values for the convergence year of
\AD\ $1091 \pm 34$, and the convergence position of RA =
\Ra{5}{34}{33}{27}, decl.\ = \dec{22}{00}{42}{52} (J2000).

\textcolor{black}{These convergence year and position estimates are not
  sensitive to the exact choice of the prior distribution for
  reasonable choices.} We emphasize that our estimate
of the convergence year is obtained using exactly the same
measurements as \citet{Nugent1998}, and the difference between our
value and the one Nugent obtained (of \AD\ $1130 \pm 16$) is due our
Bayesian estimate not being biased towards the present.  The bias in
Nugent's (and other earlier values of the convergence year) is not
large, although we note that the uncertainty in the convergence year
from our Bayesian analysis is larger than that of Nugent's estimate,
which is because our analysis takes the uncertainties in the measured
proper motions into account.

\section{Discussion}

\subsection{Expansion and acceleration of the Crab}

We found above that the fractional expansion rate of the Crab's
synchrotron bubble, \Rexp, during the period 1982 to 2012 was $0.134
\pm 0.005$~\pyear, (at mean epoch 1996.6).  This value of
\Rexp\ implies a convergence year of \AD\ $1255 \pm 27$.
\citet{Crab-expand} found \Rexp\ for epoch 1987.4 was $0.133 \pm
0.016$ and a convergence year of \AD\ $1243 \pm 92$ for the whole
nebula, which agrees well with our determination
\textcolor{black}{\citep[we note that since we used some of the same
    data as][our \Rexp\ estimates are not completely independent from
    theirs]{Crab-expand}.}

It is well known that the Crab Nebula has been accelerated since the
explosion \citep[e.g.,][]{Trimble1968, WyckoffM1977, Crab-expand,
  Nugent1998}.  The explosion epoch of 1054.5 implies that without
acceleration, \Rexp\ should be of 0.106~\pyear\ at epoch 1996.6.  Our
measured expansion rate for the synchrotron bubble of $0.134 \pm
0.005$~\pyear\ suggests therefore an accelerated expansion and the later
convergence date  of \AD\ $1255 \pm 27$.  If we assume a power-law
expansion, with $\RPWN \propto t^m$, then we can calculate that $m =
1.264 \pm 0.049$.
For a spherical pulsar wind nebula expanding into un-shocked supernova
ejecta, both theory and simulations predict an approximately power-law
expansion, with $m$ in the range of 1.1 to 1.3
\citep[e.g.][]{Chevalier1984-PWN, vdSwaluw+2001, Bucciantini+2003,
  GaenslerS2006}.  Our results are therefore consistent with the
theoretical expectations, albeit at the higher end of the range of
expected values of $m$.

We also determined the convergence year for the optical filaments to
be \AD\ $1091 \pm 34$, which implies a powerlaw exponent of $m = 1.040
\pm 0.039$.  Since we found $m = 1.264 \pm 0.049$ for the synchrotron
bubble, we can say that the synchrotron bubble is experiencing an
acceleration which is stronger than that of the optical filaments by
$3.5\sigma$.

\citet{RudieFY2008} examined the proper motion of knots in the
northern jet, at the northern extremity of the Crab, and found a
convergence date of \AD\ $1055 \pm 24$. Although this convergence date
is earlier than the biased values obtained by other workers for the
filaments in the body of the nebula, it is in fact consistent within
the combined uncertainties with the unbiased convergence date we
obtained for the optical filaments. 
\citet{RudieFY2008} conclude that the jet, unlike the filaments in the
body of the nebula, had experienced essentially no acceleration.
Since we find a lower acceleration for the bulk of the filaments, the
difference between any acceleration experienced by the jet and that
experienced by the remainder of the filaments is less clear, and we
cannot conclusively say whether or note the jet has experienced less
acceleration than the body of the nebula.

The origin of the optical filaments is generally thought to be the
following: the pulsar outflow blows a synchrotron-emitting bubble,
whose interior has high pressure but low density, into the still
freely expanding supernova ejecta.  The interface between the pulsar
bubble and the ejecta is Rayleigh-Taylor unstable
\citep{ChevalierG1975, Hester+1996, Hester2008}.  This picture is
supported by the fact that the filaments seem to only occur in a thick
shell around the exterior of the nebula, but not in the central region
\citep{Lawrence+1995, Charlebois+2010}.  The magnetohydrodynamic simulations of
\citet{PorthKK2014b} show that this instability causes ``fingers'' of
ejecta to develop, which then stream downwards into the PWN, while
``bubbles'' of synchrotron-emitting fluid move outward between them.

In this model, on average the synchrotron nebula would expand more
rapidly than the filaments.  Our observations show exactly this, and
therefore give strong support to the idea that the presently visible
optical filaments are largely the result of the Rayleigh-Taylor and
other instabilities at the interface between the synchrotron-emitting
relativistic fluid from the pulsar and the massive supernova ejecta.

In the alternate scenario of SN 1054 being a low-energy
electron-capture event \citep[e.g.][]{Smith2013, TominagaBN2013,
  Moriya+2014}, interaction of the supernova shock with the circumstellar
medium (CSM) was responsible for the luminosity of SN~1054 as well
as for the presently visible filaments, which contain largely
swept-up CSM rather than supernova ejecta.
The PWN is therefore confined by the dense wind from the super-asymptotic giant
branch progenitor of the supernova.  In this scenario one would also
expect, as we have observed, that the synchrotron nebula expands more
rapidly than the filaments.  
However, our measurement of the strong acceleration of the synchrotron
bubble is at odds with this scenario: if this bubble is confined
largely by a stellar-wind CSM, expected to have speeds of order
10~\kms, or less than 1\% of the PWN's present expansion speed, then
one would expect that PWN have been strongly decelerated since the
explosion.  The opposite seems to be true, as confirmed by our
measurement of the relatively strong acceleration of the synchrotron
bubble.  Our measurement of the acceleration of the synchrotron bubble
therefore rule out any scenario where the PWN is interacting with a
slowly moving wind.

Recently, \citet{YangC2015} \textcolor{black}{suggested} a modified
version of the canonical scenario above, where the Crab was indeed the
result of a low-energy ($\sim 10^{50}$~erg) supernova.  However,
although the energy (and mass) of the ejecta is lower than
\textcolor{black}{those of} a supernova of normal energy ($10^{51}$
erg), the synchrotron bubble is at present still interacting with the
freely expanding ejecta rather than with the CSM\@.  \citet{YangC2015}
find that for a total ejecta mass of 4.6~\Msol\ \citep[as found
  by][]{FesenSH1997}, a supernova energy of
$\lesssim 10^{50}$~erg is required, in which case the synchrotron
bubble is still interacting with the inner, flat density-profile, part
of the freely-expanding ejecta, although it is approaching point in the
ejecta density profile where the density profile becomes steep.

Our determination of the relatively strong acceleration of the
synchrotron bubble (near to the canonical value of \textcolor{black}{$r
  \propto t^{6/5}$}) is consistent with the low-energy supernova
scenario of \citet{YangC2015}, although it does not distinguish
between that scenario and that of a conventional ($E \sim
10^{51}$~erg) supernova.  In the case of a low-energy event, the
pulsar bubble might be expected, over the next few centuries, to
accelerate further once it reaches the steeper portion of the ejecta
density profile, and then decelerate as it starts to interact with the
CSM, which is moving much more slowly than the ejecta.

\section{Summary and conclusions}
\begin{trivlist}

\item{1.} By comparing our radio images from 2012 with earlier ones
  dating as far back as 1981, we conclude that the average fractional
  expansion rate of the Crab nebula's synchrotron bubble over this
  period is $\Rexp = 0.135 \pm 0.005$~\pyear.  This corresponds to a
  convergence date of \AD\ $1226 \pm 27$, or a powerlaw expansion
  since \AD\ 1054 with $\RPWN \propto t^{1.264 \pm 0.049}$.

\item{2.} We also re-examined the proper motion measurements made
  \textcolor{black}{for} the optical line-emitting filaments.  We find a
  that previous estimates of the convergence year were slightly
  biased, and \textcolor{black}{we obtain a new, un-biased Bayesian
    estimate \AD\ $1091 \pm 34$, corresponding to} a powerlaw
  expansion with $\RFIL \propto t^{1.040 \pm 0.039}$.

\item{3.} The synchrotron bubble shows significantly stronger
  acceleration than the optical line-emitting filaments. This finding
  give strong support to the idea that the presently visible optical
  filaments are largely the result of the Rayleigh-Taylor and other
  instabilities at the interface between the synchrotron-emitting
  relativistic fluid from the pulsar and the massive supernova ejecta.

\item{4.} The relatively strong acceleration of the PWN since the
  \AD\ 1054 seems to rule out scenarios where the PWN is confined
  largely by a slowly moving CSM, but rather requires that the PWN
  \textcolor{black}{be} still expanding into the freely expanding
  ejecta.  This in turn requires that the total energy in the ejecta
  is rather larger than the $\sim 5 \times 10^{49}$~erg in the
  presently visible filaments.  It therefore argues against the
  scenario where SN~1054 was an electron capture supernova producing
  $< 10^{50}$~erg.
 
\end{trivlist}

\section*{Acknowledgements}

Research at Hartebeesthoek Radio Astronomy Observatory was partly
supported by National Research Foundation (NRF) of South Africa.
Research at York University was partly supported by NSERC\@. We have
made use of NASA's Astrophysics Data System Bibliographic Services.

\bibliographystyle{mn}
\bibliography{mybib1}
\end{document}